\documentstyle[11pt]{article}
\def\ft#1#2{{\scriptstyle {#1 \over #2}}}

\def\ww3{{$W_3$}}

\def\del{\partial}
\def\p{\partial}

\def\s{\sigma}

\def\be{\begin{equation}}
\def\ee{\end{equation}}
\def\ba{\begin{eqnarray}}
\def\ea{\end{eqnarray}}

\begin{document}
\topmargin 0pt
\oddsidemargin 5mm
\begin{titlepage}
\begin{flushright}
CTP TAMU-2/95\\
SISSA 125/95/EP\\
hep-th/9511161\\
\end{flushright}
\vspace{1.5truecm}
\begin{center}
{\bf {\Large Physical States for Non-linear $SO(N)$ Superstrings}}
\vspace{1.5truecm}

{\large Z. Khviengia, H. L\"u\footnote{Supported in part by the
U.S. Department of Energy, under grant DE-FG05-91-ER40633},\ \  C.N.
Pope$^{1,}$\footnote{Supported in part by the EC Human Capital and Mobility
Programme, under contract ERBCHBGCT920176} and
E. Sezgin\footnote{Supported in part by the National Science
Foundation, under grant PHY-9411543}}
\vspace{1.1truecm}

{\small Center for Theoretical Physics, Texas A\&M University,
                College Station, TX 77843-4242}
\vspace{1.1truecm}


\end{center}
\vspace{1.0truecm}

\begin{abstract}
\vspace{1.0truecm}

We study some low-lying physical states in a superstring theory
based on the quadratically non-linear $SO(N)$--extended superconformal
algebra. In the realisation of the algebra that we use, all the
physical states are discrete, analogous to the situation in a
one-scalar bosonic string.  The BRST operator for the $N=3$ case needs
to be treated separately, and its construction is given here.

\end{abstract}

\end{titlepage}

\newpage

\pagestyle{plain}
\section{Introduction}

      Superstring theories based on  $N=1$ superconformal algebras are the
most realistic and the most commonly studied string theories so far. It is
natural, however, to look for string theories with higher worldsheet
symmetries, and to investigate their properties. If one insists on
supersymmetry and linearity of the underlying conformal algebra, the list of
possibilities is fairly short, namely the $N$--extended superconformal
algebras with $N\le 4$, and their twisted and truncated versions. However,
if one relaxes the requirement of linearity, then there exists two kinds of
superconformal algebras that go beyond $N=4$. These are the super--$W$ type
algebras, which include generators with spin higher than two, and the
higher-extended (i.e. $N>4$) superconformal algebras which are also
nonlinear, but they do not contain generators with spin higher than two
\cite{Jack}.

     String theories based on  W-type algebras are  very complicated, and so
far significant progress has been made only in the study of bosonic
$W$--strings. A nice feature of the $W$--algebras is that their realisation
allows a Minkowskian spacetime interpretation. In the case of
higher-extended superalgebras, the situation is different, in that while
they are algebraically simpler, because they do not contain higher-spin
generators, their known  realisations do not seem to allow a Minkowskian
spacetime interpretation. Instead, they seem to live on group manifolds.
Nonetheless, studies of strings on manifolds other than Minkowski spacetime
have been very useful in the past, and we expect that the situation may be
similar here.

	With these considerations in mind, in this paper we take a first
look at the properties of a string theory based on an $SO(N)$--extended
superconformal algebra, where our main interest is for $N>2$. The quantum
BRST operator for this algebra has already been constructed \cite{vn,ez}.
The nilpotency of the BRST operator imposes conditions on the central
extensions occurring in the algebra.  The critical central charge of the
energy-momentum tensor for a string theory based on $SO(N)$--extended
superconformal algebra is ${\rm dim} SO(N) + 1$. Realisations of the algebra
are not easy to come by.  So far, there exists only one realisation that
satisfies the conditions imposed on the central extensions \cite{mat}. It
makes use of a real scalar with a nonvanishing background charge, $N$ real
fermions and Kac-Moody currents of $SO(N)$ which may be constructed out of
any fields.

   In this paper, we shall look at the spectrum of physical states. An
analysis of the complete spectrum is a highly nontrivial and complicated
task which goes  beyond the scope of the present paper. Instead, we shall
investigate the spectrum of physical vertex operators at the standard ghost
sector.  We shall also look for examples of vertex operators in the
nonstandard ghost sector.

     We also investigate the case of $N=3$, which turns out to require
special treatment.  This is because the Kac-Moody level must vanish for
nilpotence of the BRST operator.  A suitable realisation of the algebra is
not known.  However, in this paper, we construct the abstract BRST operator
for this case.

\section{The $SO(N)$--Extended Superconformal Algebra and Its Realisation}

        The $SO(N)$ nonlinear superconformal algebra is generated by the
energy-momentum tensor $T(z)$, the conformal dimension 3/2 fermionic
supercurrents $G^i(z), i=1,\cdots ,N$ and the dimension 1 Kac-Moody currents
$J^a(z), a=1,\cdots ,N(N-1)/2$. In the OPE language we have \cite{Jack}
\ba
G^i(z)G^j(w)& \sim & \ {C\, \delta^{ij}\over
(z-w)^3}+ {\sigma \lambda_{ij}^a J^a(w)\over (z-w)^2} +{\ft12
\sigma \lambda_{ij}^a \partial J^a(w)\over (z-w)} +{2\delta^{ij}
T(w)\over (z-w)} +{\mu P^{ij}_{ab} (J^a J^b)(w)\over (z-w)} \ ,
\nonumber\\
J^a(z) G^i(w) &\sim &\ {-\lambda^a_{ij}\, G^j(w)\over (z-w)}
\ , \label{Nalg}\\
 J^a(z) J^b(w) &\sim &\ {-\ft12 k\epsilon\, \delta^{ab}\over (z-w)^2}+
{f^{abc} J^c(w)\over (z-w)} \ ,\nonumber
\ea
where the generators $\lambda^a_{ij}$  and the structure constants
$f_{abc}$ satisfy
\ba
&&[\lambda^a,\lambda^b] = f^{abc} \lambda^c \ ,  \qquad\qquad
{\rm tr}\, (\lambda^a\lambda^b)=- \epsilon\delta^{ab}  \ , \nonumber\\
&&\lambda^a_{ij}\lambda^{a}_{k\ell}={\epsilon\over 2} \left(\delta_{ik}
\delta_{j\ell}-
\delta_{i\ell}\delta_{jk}\right) \ , \qquad
f^{acd}f^{bdc} := -c_V \delta^{ab}=-\epsilon (N-2) \delta^{ab} \ ,\label{conv}
\ea
where $\epsilon$ is the square of the longest root (in our conventions,
$\epsilon=2$ for even $N$, and $\epsilon=4$ for odd $N$). The normal
ordering in the nonlinear term is defined by $:(JJ)(w) :\, = {1\over 2\pi
i}\oint d\zeta {J(\zeta)J(w)\over (\zeta-w)}$, and furthermore the tensor
$P^{ab}_{ij}$ is given by
\be
P^{ab}_{ij}=  (\lambda^a\lambda^b)_{ij}+(\lambda^b\lambda^a)_{ij}
              +2\delta_{ji}\delta^{ab}\ . \label{Pdef}
\ee

     The closure requirement of the OPE algebra (\ref{Nalg}) imposes the
following conditions on the parameters ocurring in the algebra \cite{Jack}:
\be
\mu = {2\over \epsilon^2 (k+N-3)}\ ,\qquad
\sigma = \mu\epsilon (2k+N-4)\ , \qquad
      C= {k\epsilon\over 2}\sigma \ .  \label{consts}
\ee
The remaining part of the algebra involves the energy momentum tensor.
It has the standard form with central extension given by
\be
c= {k (6k+N^2-10)\over 2(k+N-3)}\ . \label{cent}
\ee

  The algebra (\ref{Nalg}) can be realised in terms of a real scalar $\phi$
with a background charge $\alpha$ and $N$ real fermions $\psi^i$ as
follows \cite{mat}
\ba
T &=&  -\alpha \partial^2 \phi - \ft12 \partial \phi \partial\phi
-\ft12 \psi^i\partial \psi^i -{1\over \epsilon (k+N-3)} K^a K_a\ , \nonumber\\
G_i&=& 2i\alpha \partial\psi_i
+ i\partial \phi \psi_i - {4i \alpha\over \ell\epsilon}
\lambda^a_{ij} K_a \psi^j\ , \label{realise}\\
J^a&=& K^a+ \ft12 \lambda^a _{ij} \psi^i \psi^j\ .\nonumber
\ea
Here the currents $K^a$ obey a Kac-Moody algebra with level $\ell$, and we
must require the relations
\ba
\ell&=& k-1\ ,  \nonumber\\
\alpha^2&=&{\ell^2\over 4(k+N-3)} \ . \label{back}
\ea
The free fields $\phi$ and $\psi$ obey the OPE's:
\be
\phi(z)\phi(w)\sim\, -{\rm log}\ (z-w)\ , \qquad\qquad \psi_i(z)\psi_j(w)
\sim \delta_{ij}(z-w)^{-1}\ . \label{basicOPEs}
\ee
Note that the realisation (\ref{realise}) has a divergent coefficient when
$k+N-3=0$, in which case the extended superconformal algebra (\ref{Nalg})
has singular structure constants.  The algebra becomes degenerate since
after rescaling the fermionic currents $G^i$ to achieve non-singular
structure constants, the central extension for the fermionic currents
becomes zero.  As we shall see in the next section, for the case of $N=3$
the criticality condition precisely requires that $k=0$, and hence the BRST
analysis for this case has to be treated separately.

For $N=2$, the realisation (\ref{realise}) reduces to the one discussed
in \cite{n2}, in the special case of one complex scalar and one complex
fermion. The real part of the complex scalar corresponds to $\phi$, while
its imaginary part is used to realise the $U(1)$ current $K$. In fact, the
$N=2$ case allows a multi complex scalar and fermion realisation, for which
the physical states have been studied in \cite{n2}.

\section{The BRST Operator}

It turns out that the BRST operator for the algebra (2.1) is nilpotent
provided that \cite{vn,sch}
\be
k=6-2N\ . \label{kvalue}
\ee
Note that $k=0$ for $N=3$, in which case the algebra (\ref{Nalg}) becomes
singular.   Thus the case of $N=3$ has to be discussed separately, and will
be given at the end of this subsection.   For now we shall proceed with the
general discussion for $N\ge 4$.  With the criticality requirement of the
level number $k$ given in (\ref{kvalue}), the background charge becomes
\be
\alpha= {2N-5\over 2\sqrt {3-N}}\ .\label{acrit}
\ee
One can check that the matter contribution to the central extension is given by
\ba
c &=& ({N\over 2})+ (1+12\alpha^2)
    +\left( {\ell {\rm dim}\, G\over \ell+C_V} \right) \ , \nonumber\\
 &=&N^2-12N+26\ , \label{ccrit}
\ea
where the parantheses in the first line indicate the contributions of
$\psi^i$, $\phi$ and the $K^a$ respectively.  The parameter $\alpha$ assumes
the value given in (\ref{acrit}), and $C_V$ is defined in (\ref{conv}).
This value of $c$ is exactly cancelled by the total ghost contributions of
spin 2, $\ft32$, and 1, which is given by $(-1)^{2s+1}\, (12s^2-12s+2)$ for
each field of spin $s$.

       The BRST operator is given by
\ba
Q&=& Q_0+Q_1+Q_2+Q_3\ , \nonumber\\
Q_0&=& c\left(- \ft12 \p\phi \p\phi -\alpha \p^2 \phi
-\ft12 \psi^i\p\psi^i -{1\over \epsilon (3-N)} K^a K_a- \beta_a\p \gamma^a
-\ft32r^i\p s^i-\ft12\p r^i\, s^i \right) \ , \nonumber\\
Q_1&=& s^i\left( 2\alpha \partial\psi_i
+ \partial \phi \psi_i - {4 \alpha\over \ell\epsilon}
\lambda^a_{ij} K_a \psi^j\right)
- bs_i s^i + {2\over\epsilon}\lambda^a_{ij}
 \beta_a s_i\partial s_j\ ,
\nonumber\\
Q_2&=&\gamma_a\left( K^a+ \ft12 \lambda^a _{ij} \psi^i \psi^j
-\ft12 f_{bc}{}^a \beta^b \gamma^c +\lambda_{ij}^a\, r_i\, s_j
\right)\ , \nonumber\\
Q_3 &=& -\ft12 \mu P^{ab}_{ij}J_ar_b s^i s^j-\ft1{24}
\mu^2 P^{ab}_{ij} P^{cd}_{kl}f_{ac}{}^e
\beta_b \beta_d \beta_e s^i s^j s^k s^\ell \ .\label{BRST}
\ea
The ghost-antighost pairs $(c,b)$, $(s_i,r_i)$, $(\gamma_a, \beta_a)$
correspond to the generators $T, G^i$ and $J^a$,
respectively. They satisfy the following OPEs:
\be
 c(z)b(w)\sim (z-w)^{-1}\ , \qquad s^i(z) r_j(w)\sim \delta^i_j
(z-w)^{-1}\ , \qquad \gamma^a(z)\beta^b(w)\sim
\delta^{ab}(z-w)^{-1}\ . \label{ghOPE}
\ee

Denoting the spin-$s$ generators by $T^{(s)}$, and the corresponding ghost
fields by $(c^{(s)}, b^{(s)})$, with $s=1,\ft32,2$, we can define ghost
generators $T_{\rm gh}^{(s)}$ via the equations $[Q,b^{(s)}\}\equiv
T^{(s)}_{\rm tot} =T^{(s)}+T_{\rm gh}^{(s)}$. For linear algebras, one can
then write $Q=\sum c^{(s)}\left(\, T^{(s)}+\ft12 T^{(s)}_{\rm gh} \right)$.
However, in our case while $T_{\rm gh}$ and $J_{\rm gh}^a$ obey the same
algebra as $T$ and $J$ do, $G_{\rm gh}$ does not obey the same algebra as
$G$. Therefore, it is not as useful to write the ghostly contributions in
terms of  $T_{\rm gh}^{(s)}$. In fact, one can check they cannot all be
expressed in terms of these currents since there are seven-ghost terms in
the BRST operator.  Nonetheless, it is true here that $Q_0=c\, T_{\rm tot}$
and $Q_2=\gamma_a\, J^a_{\rm tot}$.

The problem of determining the spectrum of physical states amounts to finding
all nontrivially BRST invariant vertex operators built from the matter and
ghost fields. A physical state corresponding to a physical vertex operator
$V(z)$ is expressed as $V(0)|0>$, where $|0>$ is the $SL(2,C)$ invariant
vacuum. Denoting the modes of spin--$s$ ghost system by
$(c^{(s)}_n, b^{(s)}_n)$,
we recall that the $SL(2,C)$ ivariant vacuum has the property
$c^{(s)}_n|0>=0$ for $n \ge s$ and $b^{(s)}_n|0>=0$ for $n\ge (1-s)$.

    Like the $N=1$ NSR superstring, it is necessary to bosonise the
ghost fields $(s_i, r^i)$ for the fermionic currents $G^i$:
\be
s_i=\eta_i\, e^{\s_i}\ ,   \qquad\qquad r^i=\p \xi^i\, e^{-\s_i}\ ,
\ee
where $\s_i$ are scalar fields, and $(\eta_i, \xi^i)$ are anticommuting spin
$(1,0)$ fields. They obey the OPEs
\be
\s_i (z) \s_j(w) \sim -\delta_{ij} \, {\log}\ (z-w)\ ,
\qquad \qquad
\eta_i(z) \xi^j(w) \sim -{\delta_i^j\over z-w}\ .
\ee

In terms of the bosonised fields the BRST operator can be written as
\ba
Q&=& Q_0+Q_1+Q_2+Q_3\ , \nonumber\\
Q_0&=& c\bigg(- \ft12 \p\phi \p\phi -\alpha \p^2 \phi
-\ft12 \psi^i\p\psi^i -{1\over \epsilon (3-N)} K^a K_a-
\beta_a\p \gamma^a -b\p c
\nonumber\\
&&-\ft12 (\partial\sigma_i)^2-\partial^2\sigma_i
-\eta_i\partial\xi_i \bigg) \ , \nonumber\\
Q_1&=& \eta^i\, e^{\sigma_i} \left( 2\alpha \partial\psi_i
+ \partial \phi \psi_i - {4 \alpha\over \ell\epsilon}
\lambda^a_{ij} K_a \psi^j\right) - b\p\eta_i\eta_i e^{2\sigma_i} +
{2\over\epsilon}\lambda^a_{ij} \beta_a\eta_i e^{\sigma_i}\p
\eta_j e^{\sigma_j} \ ,
\nonumber\\
Q_2&=&\gamma_a\left( K^a+ \ft12 \lambda^a _{ij} \psi^i \psi^j
-\ft12 f_{bc}{}^a \beta^b \gamma^c-\lambda_{ij}^a\,
\p\xi_i\eta_j e^{-\sigma_i+\sigma_j}\right)\ , \nonumber\\
Q_3&=& \ft12 \mu \beta_a
\left( K^a+ \ft12 \lambda^a _{k\ell} \psi^k \psi^\ell\right)
\left( P^{ab}_{ii}\p\eta_i\eta_i e^{2\sigma_i}
+P^{ab}_{ij}\eta_i\eta_j e^{\sigma_i+\sigma_j}\right)\nonumber\\
&&-\ft1{24}
\mu^2 f_{ac}{}^e \beta_b \beta_d \beta_e\Bigg(
\ft1{12}P^{ab,cd}_{iiii}\p^3\eta_i\p^2\eta_i\p\eta_i \eta_i e^{4\sigma_i}
+\ft12 P^{ab,cd}_{iiik}\p^2\eta_i\p\eta_i\eta_i \eta_k e^{3\sigma_i+\sigma_k}
\nonumber\\
&&+P^{ab,cd}_{iikk}\p\eta_i \eta_i \p\eta_k\eta_ke^{2\sigma_i+2\sigma_k}
+P^{ab,cd}_{ijkl}\eta_i\eta_j\eta_k\eta_\ell
e^{\sigma_i+\sigma_j+\sigma_k+\sigma_\ell} \Bigg)  \ ,\label{BRST2}
\ea
where summation over repeated indices is understood, and
\be
P^{ab,cd}_{ijk\ell} :=P^{ab}_{(ij} P^{cd}_{k\ell)} \ .
\ee
It is to be understood that
an expression such as $e^{\sigma_1+\sigma_2}$ really means $:e^{\sigma_1}:
\, :e^{\sigma_2}:$, which equals $-:e^{\sigma_2}: \, :e^{\sigma_1}:$ since
both of these exponentials are fermions. Thus we have
$e^{\sigma_1+\sigma_2}= -e^{\sigma_2+\sigma_1}$ in this rather elliptical
notation.  The total energy-momentum tensor $T_{\rm tot}=\{Q, b\}$ and
total spin--1 current $J^a_{\rm tot}=\{Q, \beta^a\}$ are given
by
\ba
T_{\rm tot}&=& -\ft12(\p\phi)^2 - \alpha\p^2\phi -\ft12\psi^i\p\psi^i -
{1\over\epsilon (3-N)} K^aK^a \ ,\nonumber\\
&&- 2b\p c -\p b\, c -\beta_a\p\gamma^a
-\ft12 (\p\s_i)^2 -\p^2 \s_i - \eta_i\p \xi_i \\
J^a_{\rm tot} &=&  K^a+ \ft12 \lambda^a _{ij} \psi^i \psi^j
- f_{bc}{}^a \beta^b \gamma^c-\lambda_{ij}^a\,
\p\xi_i\eta_j e^{-\sigma_i+\sigma_j}\ .
\ea
The non-vanishing inner product for the ghost fields is
given by
\be
<0|\,\p^2c\,\p c\, c\, \gamma^1 \gamma^2\cdots \gamma^{N(N-1)/2}\,
e^{-2\sigma_1-2\sigma_2-\cdots -2\sigma_N}\, |0>=1\ . \label{innp}
\ee

    In analysing the spectrum of BRST nontrivial physical states, it is
useful to keep in mind the picture changing operators and the notion of
conjugate states. The only picture changing operators in this theory are
given by $Z^{i}=[Q,\xi^i]=-c\partial\xi^i+\cdots$. As for the conjugate
states, they are defined as follows. Given a state characterised by
$V(0)|0>$, its conjugate is given by $<0|V^\dagger$, where $V^\dagger$ is
constructed in such a way that  $<0|V^\dagger V |0>$ is proportioanl to the
inner product defined in (\ref{innp}). Checking the BRST invariance of the
conjugate states is a very convenient tool in checking the BRST nontriviality
of a given state. In order to show that a given BRST invariant state is BRST
nontrivial, it suffices to check that its conjugate state is BRST invariant.

    Before ending this subsection, we shall discuss the BRST operator for
the case of $N=3$.  The $SO(3)$-extended superconformal algebra is
generated by $J^i$ and $G^i$ for $i=1,2,3$, whose OPEs are given by
\ba
G^i(z)G^j(w) &\sim& {k-1 \over (z-w)^3} +{2(1-k)\over k} {\varepsilon^{ijk}
J^k\over (z-w)^2} + {2 \delta^{ij} T + {1-k\over k} \p J + {1\over k}
(J^iJ^j + J^j J^i) \over z-w}\ ,\nonumber\\
J^i(z) G^j(w) &\sim& {\varepsilon^{ijk} G^k \over z-w}\ ,\qquad
J^i(z)J^j(w) \sim {-\ft12 k\over (z-w)^2} + {\varepsilon^{ijk} J^k
\over z-w}\ ,\label{N3alg1}
\ea
together with the energy-momentum tensor $T$ with central charge
$c=(-1+3k)/2$.   The criticality condition $c^{\rm tot}=c^{\rm mat} + c^{\rm
gh}=0$ implies that $k=-\ft13$; however, the condition that $J^i_{\rm tot}$
have zero central extension requires that $k=0$.   Thus one cannot build
a nilpotent BRST operator for the algebra (\ref{N3alg1}).  However, when
$k=0$ the algebra (\ref{N3alg1}) becomes singular.    We can rescale the
fermionic currents, $G^i \longrightarrow G^i/\sqrt{k}$, to obtain a
contracted algebra for $k=0$.  In terms of the rescaled currents, the algebra
(\ref{N3alg1}) becomes:
\ba
G^i(z) G^j(w) &\sim& {2\varepsilon^{ijk} J^{k} \over (z-w)^2} +
{\varepsilon^{ijk}\p J^k + J^i J^j + J^j J^i \over z-w}\nonumber\\
J^i(z) G^j(w) &\sim& {\varepsilon^{ijk} G^{k}\over z-w}\ ,\qquad
J^i(z) J^j(w) \sim {\varepsilon^{ijk} J^k\over (z-w)}\ ,\label{N3alg2}
\ea
together with the energy-momentum tensor $T$ with an
arbitrary value of central charge, which can be set to the critical value
$c=-1$.  We find that the BRST operator is given by
\ba
Q&=& c(T - b\del c + \ft32 r^i\p s^i + \ft12 \p r^i\, s^i -\beta^i\p \gamma^i)
+s^i G^i + \gamma^i J^i\nonumber\\
&& + \varepsilon^{ijk}\gamma^i r^j s^k -\ft12
\varepsilon^{ijk}\gamma^i\beta^j\gamma^k - J^i\beta^j s_i s_j\ .
\label{N3brst}
\ea

\section{Physical States}

Starting from the BRST operator (\ref{BRST2})
for the $SO(N)$-extended superconformal algebra for $N\ge
4$, we shall now study some of the physical states in this section.
The simplest BRST invariant operator is clearly the unit operator,
corresponding to the $SL(2,C)$ vacuum state.  The next natural physical
operator to
consider is the one which corresponds to the lowest energy state.
It is given by
\be
V_0= ce^{-\sigma_1-\sigma_2\cdots -\sigma_N}\xi^{A}\, \Phi_A\, e^{p\phi}
\ ,\label{v0}
\ee
where $A$ labels an arbitrary representation of $SO(N)$, $\xi^A$ is a
polarisation vector, and $\Phi_A$ is the primary field under the Sugawara
energy-momentum tensor $T_{SGW}(z)=-{1\over \epsilon (3-N)} K^aK_a$.
The field $\Phi_A$ satisfies the following OPEs:
\ba
K^a(z)\Phi_A(w) &=&{-\tau^a_{AB} \Phi^B (w) \over (z-w)}\ ,
\\
T_{SGW}(z)\Phi_A (w)&=& {C_R \over \epsilon (3-N)} {\Phi_A(w) \over (z-w)^2}+
{\p \Phi_A\over (z-w)}\ ,
\ea
where  $\tau^a_{AB}$ are the $SO(N)$ generators in the representation
R and $C_R$ is the eigenvalue of the second  Casimir in this representation:
$(\tau^a\tau^a)_{AB}=-C_R\delta_{AB}$. In particular, for the $r$--th
rank totally
antisymmetric representation we have  $C_R (r)={\epsilon\over 2}r(N-r)$.

    The physical operator $V_0$ (\ref{v0}) has standard ghost structure, in
that it is built with the standard ghost vacuum vertex operator
$ce^{-\sigma_1-\sigma_2\cdots -\sigma_N}$. For such states, the
physical-state condition can be summarised as follows:
\ba
L_0\big|{\rm Phys}\big\rangle_{\rm mat}&=&(1-\ft{N}2)\big|{\rm Phys}
\big\rangle_{\rm mat}\ , \qquad
L_n\big|{\rm Phys}\big\rangle_{\rm mat}=0\, \quad n\ge 1\nonumber\\
G^i_{n+\ft12}\big|{\rm Phys}\big\rangle_{\rm mat}&=&0\ , \qquad
J^a_n\big|{\rm Phys}\big\rangle_{\rm mat}=0\ , \quad n\ge 0\ ,\label{physcon}
\ea
where the currents are constructed from matter only.  The matter part of the
physical operator $V_0$, being tachyonic, is already annihilated by the
positive modes of the currents.  However, the $J_0$ condition implies that
$\Phi^A$ has to be a singlet under the Kac-Moody currents $K_a$, and the
remaining mass-shell condition determines that the momentum for the
scalar field $\phi$ is given by
\be
p_\pm = {1\over 2\sqrt{3-N}} \Big(- 2N+5 \pm 1\Big)\ .
\ee
In checking this, and in further calculations, it is useful to note the
dimension formulae $\Delta \left(e^{p\phi}\right)=-\ft12 p(p+2\alpha)$, and
$\Delta \left(e^{q\sigma}\right)=-\ft12 q(q+2)$.  Furthermore, for a scalar
field $\phi$ satisfying the OPE given in (\ref{basicOPEs}), the following OPE
holds: $e^{a\phi(z)} \, e^{b\phi(w)}=(z-w)^{-ab}\, e^{a\phi(z)+b\phi(w)}$.

     Higher-level physical states with standard ghost structure can be
obtained by acting with excitations of the basic fields on the lowest-energy
state $V_0$.  Thus the matter part of the vertex operators take the form
\be
V= R^A(\del\phi, \psi^i, K^a)\, \Phi_A\, e^{p\phi}\ ,\label{physgen}
\ee
where $R^A$ is a polynomial in the basic fields $\del \phi, \psi^i$ and
$K^a$.   $R^A$ can be characterised by its conformal dimension, {\it
i.e.~}its level number.   For a physical operator with level number $n$, the
mass-shell condition implies that
\be
- {p^2 \over 2} -\alpha p+{C_R \over \epsilon(3-N)} -1 + {N \over 2}+n =0 \ .
\label{masscon}
\ee

     To obtain the explicit form of a higher level physical operator
(\ref{physgen}), it is necessary to solve the physical-state conditions
(\ref{physcon}), which become very complicated with increasing
level number.  However, we can discuss certain general features of the
physical states with standard ghost structure.   Since the basic fields and
the currents have a one-to-one correspondence, {\it i.e.}
\be
T\longleftrightarrow \p\phi\ , \qquad
G^{i}\longleftrightarrow \psi^i\ ,\qquad
J^a  \longleftrightarrow K^a\ ,
\ee
it follows that the matter excitations $R^A(\del\phi, \psi^i, K^a)$ can be
re-expressed as $R^A(T_n, G^i_m, J^a_k)$.  Physical states of this form are
BRST trivial.   Thus we expect that all excited physical states with
standard ghost structure are BRST trivial.  Such a phenomenon also occurs in
the one-scalar string theory and the two-scalar $W$ string, where all the
higher-level physical states with standard ghost structure are trivial.

However, this does not imply that the BRST cohomology of the system is
simple.  In fact it can have a very rich structure, since there can be many
physical states with non-standard ghost structure.   Studying the physical
states with standard ghost structure, namely the BRST trivial states, can
unveil the physical states with non-standard ghost structure.    To see
this, we note that a higher-level physical state (\ref{physgen}) can be
written as $Q\chi$ for generic on-shell momentum since it is BRST trivial.
However, if for certain a special value of on-shell momentum the state
becomes zero, then it implies that the operator $\chi$ becomes BRST
invariant for this special value of momentum.  Obviously the operator $\chi$
has a non-standard ghost structure, and it corresponds to a BRST-non-trivial
physical state.

      First we shall look at the physical states with level number $\ft12$,
which have the form
\be
V=\xi^{i\, A} G^{i}_{-\ft12} \Phi_A\, e^{p\phi} =
-\xi^{i\, A}\Big( p\, \psi_i\, \Phi_A + {2\over\sqrt{3-N}}\,
  \lambda^c_{ij}\, \tau^c_{AB}\, \psi_j\, \Phi_B\Big) e^{p\phi} \ .
\ee
The only non-trivial physical-state conditions are the $J_0^a$ and $L_0$
conditions.   The former implies that $\Phi_A$ is a singlet under the
Kac-Moody currents $K^a$; the latter implies that the momentum $p$ satisfies
the mass-shell condition (\ref{masscon}) with $C_R=0$.   This BRST trivial
state will not vanish for any on-shell momentum, and hence we do not expect
that there exists a BRST non-trivial state in the lower ghost number.

     At level $n=1$, we consider $SO(N)$--singlet physical states of
standard ghost structure, with matter-dependent factors of the form
\be
V=(\lambda^{a}_{ij} G^{i}_{-\ft12} G^{j}_{-\ft12} +
x J^a_{-1}) \Phi_a e^{p\phi}\ ,
\ee
where $x$ is a constant.  It is straightforward to show that these
states can be rewritten as
\be
V=\Big[\Big(x - {2(q+N-2)\over 3-N} \Big) K^a +
    \ft12 \Big(x + {2(q+N-2)^2\over 3-N} \Big) \lambda^a_{ij}\,
\psi_i\, \psi_j \Big]\, \Phi_a e^{p\phi}\ ,
\ee
where $q \equiv p\,  {\sqrt{ 3-N}}$.  The physical-state conditions
(\ref{physcon}) then require that the mass-shell condition
\be
(q+N-1)(q+N-4)=0 \label{mmass}
\ee
be satisfied, and then also give rise to a polarisation condition which
determines the constant $x$:
\be
x={(q+N-2)(q+N-3)\over N-3}\ .
\ee
Thus it is easy to see that if the momentum satisfies the condition
$q=1-N$, corresponding to the first root of (\ref{mmass}), the physical
state described by $V$ vanishes identically.  This is the situation
where a physical state that is generically null becomes identically
zero at a special value of the on-shell momentum.  It signals the
occurrence of a BRST non-trivial physical operator at ghost-number
one less than that of the standard physical states, since at this
momentum the null operator $Q\chi$ vanishes, implying that $\chi$ itself
becomes physical.

     For the case of $N=2$, we can easily find the explicit form
of the physical operator $\chi$.  It is given by
\be
\chi=c\, \Big( \psi_1\, \del\xi_2\, e^{-\sigma_1-2\sigma_2} +
     \psi_2\, \del\xi_1 \, e^{-2\sigma_1-\sigma_2} +2 \beta\,
     e^{-\sigma_1-\sigma_2} \Big) e^{-\phi}\ .
\ee
One can check, by constructing the operator conjugate to this,
that $\chi$ is BRST non-trivial.  For $N>3$, the analogous operator
is given by
\be
\chi=c\,\Big( \psi_i\,\del\xi_j\, e^{-\sigma_j-\Sigma\sigma}\, \lambda_{ij}^a
       +{2\over \sqrt{3-N}} \beta^a\, e^{-\Sigma\sigma} \Big)
\Phi_a e^{p\phi}\ ,
\ee
where $\exp(-\sum\sigma)=\exp(-\sigma_1-\cdots- \sigma_N)$ and
$\exp(-\sigma_j-\sum\sigma)=
\exp(-\sigma_1-\sigma_2-\cdots-2\sigma_j-\cdots-\sigma_N)$.  The momentum
$p=q/\sqrt{3-N}$ is given by $q=1-N$.

\section{Comments}

     With the realisation (\ref{realise}) for the $SO(N)$--extended
superconformal algebra, one expects that in the sector of physical states
with standard ghost structure, there should be no states with excitations of
the matter fields, and that the only non-trivial physical state should be
the tachyon.  This is because the number of matter fields is the same as the
number of constraints, and thus there are no transverse spacetime
directions.  In this paper we have seen that this is indeed true, but that
in addition there are further physical states with non-standard ghost
structure; {\it i.e.}\ with ghost as well as matter excitations.
Multi-scalar realisations, which might allow the possibility of obtaining a
spacetime interpretation, are unknown for the cases $N>2$. On the other
hand, for $N\le 2$, where the algebras are linear, multi-scalar realisations
are well known, and they indeed lead to string theories with target
spacetimes that allow for transverse excitations.  Although we have only
exhibited a small number of physical states in this paper, we expect that
there will exist infinite numbers of such states at arbitrary ghost numbers,
in much the same way as one finds in the one-scalar bosonic string.

     The case of $N=3$ is special, because the Kac-Moody level required by
nilpotency of the BRST operator is zero, which conflicts in the full
$N=3$ algebra with the value of the central charge that is required by
nilpotency.  This conflict can be circumvented by performing a rescaling
of the fermionic generators $G^i$ that becomes singular in the limit
where the Kac-Moody level tends to zero.  By this means, we have been
able to construct a nilpotent BRST operator for a contraction of the
full $N=3$ algebra.

\section*{Acknowledgement}
C.N.P. is grateful to SISSA, Trieste, for hospitality during the completion
of this work.


\begin{thebibliography}{20}

\bibitem{Jack} V.G. Knizhnik, Theor. Math. Phys. {\bf 66} (1986) 68;
    M. Bershadsky, Phys. Lett. {\bf B174} (1986) 285;
    E.S. Fradkin and V. Y. Linetsky, Phys. Lett. {\bf B282} (1992)
         352;{\bf B291} (1992) 71; Phys. Lett. {\bf B275} (1992) 345.
\bibitem{vn} K. Schoutens, A. Sevrin and P. van Nieuwenhuizen, Commun. Math.
           Phys. {\bf 124} (1989) 87.
\bibitem{ez} Z. Khviengia and E. Sezgin,  preprint, CTP TAMU-27/93.
\bibitem{sch} K. Schoutens, Nucl. Phys. {\bf B314} (1989) 519.
\bibitem{mat} P. Matthieu, Phys. Lett. {\bf B218} (1989) 185.
\bibitem{n2} H. L\"u, C.N. Pope, X.J. Wang and K.W.Xu, {\em Phys. Lett.}
{\bf B284}, (1992) 268.

\end{thebibliography}
\end{document}